\documentclass[article]{jss}

\usepackage{orcidlink,thumbpdf,lmodern}

\usepackage{framed}

\usepackage{amsmath}
\usepackage{cleveref}
\usepackage{subcaption}

\newcommand{\KM}{Kaplan--Meier} 

\author{Jeffrey Roskes~\orcidlink{0000-0001-8761-0490}\\Johns Hopkins University} 
\Plainauthor{Jeffrey Roskes}

\title{\pkg{KoMbine}: Propagating Statistical and Systematic Errors to \KM{} Curves}
\Plaintitle{KoMbine: Propagating Statistical and Systematic Errors to \KM{} Curves}
\Shorttitle{\pkg{KoMbine}: Propagating Errors to \KM{} Curves}

\Abstract{
  \KM{} curves are widely used in medical research to evaluate the performance of biomarkers and predict patient outcomes. These curves are often shown without error bands, and even when error bands are provided, they typically only account for the statistical uncertainty resulting from the finite number of patients in the study. In reality, other sources of uncertainty affect the measurements as well. As datasets grow, the statistical uncertainty on the number of patients no longer dominates the overall uncertainty, and other uncertainties are increasingly important to model. The \pkg{KoMbine} package, developed based on procedures used in particle physics, provides the first method to propagate both statistical and systematic uncertainties through the \KM{} curve estimation processes.
}

\Keywords{kaplan--meier curve, uncertainty} 
\Plainkeywords{kaplan-meier curve, uncertainty}

\Address{
  Jeffrey Roskes\\
  Institute for Data Intensive Engineering and Science\\
  \emph{and}\\
  Department of Physics and Astronomy\\
  Johns Hopkins University\\
  3400 N. Charles St\\
  Baltimore, MD 21218\\
  E-mail: \email{jroskes1@jhu.edu}
}

\begin{document}



\section{Introduction}\label{sec:intro}

\KM{} curves are widely used in medical research to predict patient outcomes by giving the survival probability as a function of time. The curves are constructed empirically based on observed outcomes in a cohort of patients. They can be used to evaluate the performance of biomarkers, by stratifying patients into groups based on the biomarker value and plotting separate \KM{} curves for each group, or to compare the effectiveness of different treatments. By examining the \KM{} curves, we can also predict how long a future patient is likely to survive.

\KM{} curves are often shown without error bands, or with error bands that only take into account the statistical uncertainty resulting from the finite number of patients in the study. In this paper, we present \pkg{KoMbine}, the first package that can calculate both statistical and systematic uncertainties on \KM{} curves.

The maximum likelihood techniques and modeling of systematic uncertainties used in \pkg{KoMbine} are based on the \pkg{Combine} package \citep{CAT-23-001}, which was developed by the CMS Collaboration in particle physics\@. \pkg{Combine}, built on the \pkg{RooFit} framework \citep{RooFit} and the \pkg{Minuit} optimizer \citep{minuit}, is used for high-precision measurements of particle properties, which must account for tens or even hundreds of systematic uncertainties, and was used for such high-profile measurements as the discovery of the Higgs boson \citep{HIG-12-028} and subsequent measurements of its properties.

No particle physics measurement is reported without a comprehensive assessment of the uncertainties that affect the measurement. This rigorous approach has led to a higher degree of reproducibility in particle physics results than in many other fields, including medical research.

The \pkg{KoMbine} package handles the statistical uncertainty on the number of patients in a more precise way than conventional methods and incorporates patient-wise uncertainties into the calculation. The statistical and patient-wise uncertainties can also be plotted separately, allowing the user to see which uncertainties dominate, which in turn can help to plan the next iteration of the experiment.

\pkg{KoMbine} is available from PyPI via \code{pip install kobminekm} and from GitHub at \url{https://github.com/AstroPathJHU/KoMbine}.

\section{Background}\label{sec:background}

\subsection{\KM{} curves}

A \KM{} curve illustrates how long a patient is expected to survive after a certain starting point. By definition, the patient is alive at the start, so the curve starts with a survival probability of \(1.0\) at time \(0\), which represents, for example, the time of treatment. The survival probability decreases as time progresses, representing patient deaths. The survival time is measured relative to the starting point, which may be a different actual time for each patient.

\KM{} curves can also be used for events of interest other than death, such as disease progression or relapse. The curve represents the probability of the patient lasting for a length of time without experiencing that event.

Sometimes, a patient drops out of a study without experiencing the event of interest; for example, the study may end without the patient experiencing the event, or the patient may die from other causes unrelated to the disease. Such a patient is said to have been ``censored.'' The patient is included in the \KM{} curve until the point of censoring. After the patient is censored, they are no longer considered ``at risk,'' and the future survival probability is calculated based on the remaining patients.

\subsection{Uncertainties on \KM{} curves}

Various popular packages, such as \pkg{survival} \citep{survival-package} in \proglang{R} \citep{R} or \pkg{lifelines} \citep{lifelines} in \proglang{Python}, use Greenwood confidence intervals to provide error bands that account for the statistical uncertainty resulting from the finite number of patients in the study~\citep{GreenwoodNotes,Greenwood}. (To be more precise, \pkg{lifelines} uses the exponential Greenwood confidence intervals, while \pkg{survival} allows several different methods, all variations of Greenwood's.)

Another statistical metric often applied to \KM{} curves is the \(p\)~value for rejecting the null hypothesis that the two curves are identical. This \(p\)~value, as traditionally calculated, also only takes into account the statistical uncertainty from the finite number of patients.

Other uncertainties that affect the \KM{} curve will typically be patient-wise uncertainties, which \emph{move} the patient from one \KM{} curve to another. For example, a biomarker study might classify patients based on the density of a particular cell type. A \KM{} plot could be constructed with two curves, for patients with a high and low density of the cell type. However, the cell type is subject to Poisson noise as well as systematic errors in the measurement. Similar errors apply to \KM{} plots that stratify patients by gene expression levels or other continuous variables.

The traditional calculations also make certain approximations, in particular approximating the binomial distribution by a normal distribution. These approximations are most valid for large numbers of patients. The commercial software \pkg{StatXact} \citep{StatXact} does provide an exact method for calculating the log-rank \(p\)~value for two \KM{} curves, but does not provide confidence intervals for a single \KM{} curve, nor does it account for patient-wise uncertainties.

\section{Methodology}\label{sec:methodology}

\pkg{KoMbine} uses a log-likelihood function to compute the confidence interval for each point on the \KM{} curve. The log likelihood is a function of the survival probability \(S\) as well as other parameters. We maximize the likelihood over those other parameters to find the ``likelihood scan'' as a function of \(S\) alone. The points of maximum likelihood form the \KM{} curve, and the confidence intervals are determined by the points where the log likelihood is below a threshold.

\subsection{Notation}\label{sec:notation}

We denote the inclusion of patient \(j\) by \(a_j\), which is \(1\) if the patient is included and \(0\) if the patient is excluded. The numbers of patients who were at risk and who died at time \(t_i\) are:
\begin{align}
r_i &= \sum_{t_j \geq t_i} a_j, \\
d_i &= \sum_{\substack{t_j = t_i \\ \neg c_j}} a_j,
\end{align}
where \(c_j\) indicates whether the patient was censored or not. We will allow \(a_j\) to vary so that \(r_i\) and \(d_i\) may differ from their nominal values \(\hat{r}_i\) and \(\hat{d}_i\).

The survival probability at time \(t_n\) is given by the \KM{} formula:
\begin{equation}
S_n = \prod_{i=1}^{n} p_i^\text{s}, \label{eq:km-survival}
\end{equation}
where \(p_i^\text{s}\) is the probability that a patient who has survived until time point \(t_i\) continues to live. The nominal probability is calculated from the observed survival probabilities:
\begin{equation}
\hat{S}_n = \prod_{i=1}^{n} \left(1 - \frac{\hat{d}_i}{\hat{r}_i}\right). \label{eq:km-nominal}
\end{equation}

\subsection{Greenwood confidence intervals}\label{sec:greenwood}

As mentioned earlier, most \KM{} curve packages provide confidence intervals using some variation of Greenwood's formula. At confidence level \(1-\alpha\), Greenwood's interval is given by
\begin{equation}
\label{eq:greenwood}
S(t)\in\hat{S}(t) \pm z_{\alpha/2} \sqrt{{\hat{S}(t)}^2{\sum_{i=1}^{n} \frac{d_i}{r_i(r_i-d_i)}}},
\end{equation}
where \(\hat{S}\) is the nominal value of the survival function, given by \cref{eq:km-nominal}. The exponential Greenwood interval is given by
\begin{equation}
\label{eq:exponential-greenwood}
\log{\left(-\log S(t)\right)} \in \log{\left(-\log \hat{S}(t)\right)} \pm z_{\alpha/2} \sqrt{\frac{1}{{\left(\log \hat{S}(t)\right)}^2} \sum_{i=1}^{n} \frac{d_i}{r_i(r_i-d_i)}}.
\end{equation}

\citet{GreenwoodNotes} gives the derivation of these formulas, which makes certain approximations: the ``delta-method'' approximation, which requires that the true and observed probabilities are close enough that \(\log\left(-\log S(t)\right)\) can be treated as approximately linear, and the approximation of the binomial distribution by a normal distribution.

With large numbers of patients, these approximations hold, but their validity decreases for smaller cohort sizes. Our log likelihood method avoids these approximations.

\subsection{Log likelihood implementation}

For clarity, we will distinguish the two types of uncertainty: the statistical uncertainty from the finite number of patients (which we will call the ``binomial uncertainty'') and the patient-wise uncertainty.

\subsubsection{Binomial uncertainty}\label{sec:binomial-uncertainty}

The binomial contribution to the \KM{} curve is computed from the number of patients at risk and the number of patients who died at each time point:
\begin{equation}
\mathcal{L}_{\text{binomial}}(p_i^\text{s}, a_j) = \prod_{i=1}^{n} \binom{r_i}{d_i} {\left(p_i^\text{d}\right)}^{d_i} {\left(p_i^\text{s}\right)}^{r_i-d_i},
\end{equation}
where \(p_i^\text{d} = 1 - p_i^\text{s}\). The negative log likelihood (NLL) is
\begin{equation}
-\log \mathcal{L}_{\text{binomial}}(p_i^\text{s}, a_j) = -\left[\sum_{i=1}^{n} \left( \log\binom{r_i}{d_i} + d_i \log p_i^\text{d} + (r_i-d_i) \log p_i^\text{s} \right)\right]. \label{eq:binomial-nll}
\end{equation}

We are not interested in the individual \(p_i^\text{s}\) values, but rather the overall survival probability \(S_n\). We therefore use \cref{eq:km-survival} as a constraint when we minimize the NLL\@.

\subsubsection{Patient-wise uncertainty}\label{sec:patient-wise-uncertainty}

We now consider that the patients themselves may be subject to uncertainty. Some patients who were nominally included in the \KM{} curve may actually be excluded or vice versa.

The classic case when this happens is when dividing the patients into groups based on a continuous parameter, such as gene expression or the density of a certain cell type. The parameter measurement is subject to uncertainty, which must be considered when determining the overall uncertainty on the \KM{} curve. (This is in contrast to \KM{} curves based on discrete categories with no error, such as treatment vs.\ placebo, where patient-wise uncertainty likely does not enter.)

We first precompute the NLL penalty for each patient to be included in the \KM{} curve, which is negative if the patient is nominally included and positive if the patient is nominally excluded. The penalty depends on the functional form of the patient's parameter's probability density function, and it is calculated as the difference between the minimum NLL for the parameter to be within the range to be included in the curve and the minimum NLL for it to be outside the range.

\pkg{KoMbine} allows several possible probability distributions for the patient parameters, including a delta function, Poisson distribution, a Poisson distribution divided by a fixed area (to represent, for instance, a density of cells), and a ratio of two Poisson distributions. Any of these can be multiplied by systematic uncertainties following the log-normal distribution, a common choice used to represent multiplicative uncertainties.

For the purpose of the math here, all that is important is that the penalty is a fixed number per patient, which we call \(-\log \mathcal{L}_j^{\text{patient}}\). We compute the NLL as
\begin{equation}
-\log \mathcal{L}_{\text{patient}}(a_j) = -\sum_{j=1}^{m} a_j\log \mathcal{L}_j^{\text{patient}}. \label{eq:patient-nll}
\end{equation}

\subsubsection{Combining the uncertainties}\label{sec:combining-uncertainties}

The total NLL is the sum of the binomial NLL and the patient-wise NLL\@:
\begin{equation}
-\log \mathcal{L}_{\text{total}}(p_i^\text{s}, a_j) = -\log \mathcal{L}_{\text{binomial}}(p_i^\text{s}, a_j) - \log \mathcal{L}_{\text{patient}}(a_j). \label{eq:total-nll}
\end{equation}

We minimize the overall NLL (\ref{eq:total-nll}) over the \(p_i^\text{s}\) and \(a_j\), with the constraint (\ref{eq:km-survival}). This is a mixed integer nonlinear programming (MINLP) problem.

\subsection{Isolating the binomial and patient-wise uncertainties}\label{sec:isolating-uncertainties}

Although the total uncertainty on the \KM{} curve includes both the binomial and the patient-wise uncertainty, it is instructive to look at them separately. Determining which uncertainty dominates can help to understand the next steps in the analysis: should we try to enroll more patients or should we improve the biomarker measurements?

\subsubsection{Binomial uncertainty}

To examine the binomial uncertainty, we can fix the parameters to their nominal values, fixing \(a_j=\hat{a}_j=1\) for the patients who are nominally included and \(0\) for those who are nominally excluded. These results are approximately equivalent to the Greenwood confidence intervals used in other packages. As will be illustrated in \cref{sec:compare-to-greenwood}, our method is somewhat more precise, but is also computationally slower.

\subsubsection{Patient-wise uncertainty}

Isolating the patient-wise uncertainty is somewhat nontrivial. We cannot allow the survival probabilities \(p_i^\text{s}\) to vary freely, as we would need the binomial term to constrain them. Instead, we fix
\begin{equation}
p_i^\text{s}(a_j) = 1 - \frac{d_i(a_j)}{r_i(a_j)}.
\end{equation}

With this definition, \(S\) can only take on certain fixed, discrete values. Furthermore, depending on the order that patients were censored or died, values of \(S\) that are close together may actually only be reachable by very different combinations of patients included in the analysis, with correspondingly different NLL values. A scatter plot of \(-\log\mathcal{L}\) as a function of \(S\) would be useless.

Instead, we use a continuous variable \(S'\). We calculate \(\hat{S}\) using the nominal values of the patient parameters, as in \cref{eq:km-nominal}. Then, for any value of \(S'\), we minimize the NLL over \(a_j\) with the constraint that \(S\) is at least as far from \(\hat{S}\) as \(S'\):
\begin{align}
S \ge S' & \text{ if } S' > \hat{S} \\
S \le S' & \text{ if } S' < \hat{S}
\end{align}
This gives an interpretable visualization of the patient-wise uncertainty: we obtain the NLL that the survival probability is at least as far from the nominal probability as \(S'\). By construction, \(-\log\mathcal{L}(S')\) has a single minimum (either a single point or, if some patients are exactly on the boundary, a flat region) at \(\hat{S}\) and increases as \(S'\) moves away in either direction.

\subsection[p~values]{\(p\)~values}

\pkg{KoMbine} also provides a method to compute the \(p\)~value to reject the null hypothesis that two survival curves are identical. We use a log-rank approach, similar to the conventional approach used in \pkg{lifelines} and similar packages. However, as in the case of the \KM{} error bars, our likelihood modeling does not rely on approximating the binomial distribution as normal, and we also include the patient-wise errors.

We have two curves, labeled \(k=0\) and \(k=1\), and we add this index to all relevant quantities. For example, each patient may be in one of the curves, and this inclusion is given by \(a_{jk}\). Each patient can be in at most one of the two curves, but we also allow for a patient to be in neither. (This is useful, for example, when stratifying patients into low, middle, and high categories based on a parameter\@. \(p\)~values are computed pairwise between the curves, and in each pairwise comparison, some patients are in neither curve.)  Each patient has a positive or negative penalty, \(-\log\mathcal{L}_{jk}^\text{patient}\), to be included in each curve.

Each curve has its probabilities of dying \(p^\text{d}_{ik}\). In the log-rank test, we take the proportional hazard model,
\begin{equation}
p^\text{d}_{i0} = H p^\text{d}_{i1},
\label{eq:logrank-hazard}
\end{equation}
for a constant hazard ratio \(H\) across all time points. The null hypothesis is that \(H=1\), meaning that the two curves are identical. The alternative hypothesis allows \(H\) to float.

In the conventional approach, we define the residual for each time point as
\begin{align}
\begin{aligned}
U_i&=d_{i1} - d_{i1}^\text{exp} \\
&=d_{i1} - (d_{i0} + d_{i1})\frac{n_{i1}}{n_{i0}+n_{i1}}.
\end{aligned}
\end{align}
By symmetry, if we swapped the two curves, \(U_i\rightarrow-U_i\). Summing the residuals and dividing by the variance gives us the test statistic,
\begin{align}
U&=\sum_i U_i \\
Q&=\frac{U^2}{\text{Var}(U)},
\end{align}
which follows a \(\chi^2\) distribution with one degree of freedom~\citep[pp.~45--47]{mantel1966evaluation,peto1972asymptotically,appliedsurvivalanalysis}.

In the likelihood approach, we instead compute the full NLL, similar to \cref{eq:total-nll}. We have two separate patient-wise NLL penalties, one for each curve. The probability distribution representing the relationship between the curves is modeled with the Cox partial likelihood distribution with \(\beta=\log{H}\), which reduces to the hypergeometric distribution under the null hypothesis when \(H=1\)~\citep{cox1972regression,breslow1974covariance}:
\begin{align}
-\log \mathcal{L}_{\text{Cox}}(a_{jk}, H)&=-\sum_i\left[\log\binom{r_{i0}}{d_{i0}}+\log\binom{r_{i1}}{d_{i1}}+d_{i1}\log H-\log Z_j(H)\right] \label{eq:cox-nll-exact}\\
&\approx-\sum_i \left[d_{i1} \log H - d_i \log\left(r_{i0} + H r_{i1}\right)\right] \label{eq:cox-nll-breslow}
\end{align}
\Cref{eq:cox-nll-exact} gives the exact formula, with the normalization factor \(Z_j(H)\) given by
\begin{equation}
Z_j(H) = \sum_{m}\binom{r_{i0}}{m}\binom{r_{i1}}{d_{i0}+d_{i1}-m}H^m,
\label{eq:cox-normalization}
\end{equation}
where the sum is over all feasible values of \(m\) that allow both binomial coefficients to be finite.

\Cref{eq:cox-nll-breslow} applies the Breslow approximation~\citep{breslow1974covariance}. Both give identical results when there no two patients die at the same time. Since our biomedical analyses typically do not involve many patients, ties are rare, and we therefore use \cref{eq:cox-nll-breslow} in \pkg{KoMbine}. We explore the effect of this approximation in \cref{sec:compare-to-conventional-p-value}.

The total NLL is then given by
\begin{equation}
-\log \mathcal{L}_{\text{total}}(a_{jk}, H) = -\sum_{k}\log\mathcal{L}_\text{Cox}(H) - \sum_{k}\log\mathcal{L}_k^{\text{patient-wise}}(a_{jk}).
\label{eq:p-value-nll}
\end{equation}

We evaluate the NLL for the null hypothesis by minimizing \cref{eq:p-value-nll} over \(a_{jk}\) with the constraint \(H=1\). For the alternate hypothesis, we minimize \cref{eq:p-value-nll} over both \(a_{jk}\) and \(H\). The difference between the two NLLs is used to compute the \(p\)~value using a \(\chi^2\) distribution with one degree of freedom.

We also allow fixing \(a_{jk}\) and using only the Cox penalty. Because floating \(a_{jk}\) can decrease the NLL for both the null and alternate hypotheses, the \(p\)~value may be either larger or smaller than the \(p\)~value with fixed \(a_{jk}\). \Cref{sec:compare-to-conventional-p-value} compares our Cox-only \(p\)~value to the conventional log-rank test.

\section{Implementation}

\pkg{KoMbine} is implemented in \proglang{Python} and uses \pkg{Gurobi} \citep{gurobi} to minimize the NLL for a given value of \(S\).

We first minimize \(-\log \mathcal{L}_{\text{total}}\) without the constraint on \(S\) (\ref{eq:km-survival}) to find the maximum likelihood estimate and the best value of \(S\)\@. \code{scipy.optimize.brentq} \citep{brentq,scipy} is then used to compute the confidence intervals, bounded by the values of \(S\) where the NLL equals a threshold value (1.0 for 68\% CL and 3.84 for 95\% CL).

For the patient-wise-only error, there is no need to minimize the NLL over \(S'\), as by construction it minimizes at \(S'=S_\text{nominal}\). To find the confidence intervals, we use a simple binary search that only looks at the possible discrete values of \(S\): intermediate values of \(S'\) cannot return anything different.

To compute the \(p\)~value, \pkg{KoMbine} similarly uses \pkg{Gurobi} to minimize the NLL for the null and alternate hypotheses. The \(p\)~value is computed from the difference between the NLLs, which follows a \(\chi^2\) distribution, using standard functions in \pkg{scipy}.

\subsection{Collapsing consecutive deaths}\label{sec:collapsing-consecutive-deaths}

The binomial term in the NLL [\cref{eq:binomial-nll}] requires a sum over all time points where at least one patient died, which can be slow when there are many such time points. A mathematical trick allows us to collapse multiple time points into one, as long as no patients are censored in between, without changing the results.

Take the case of two consecutive time points, \(t_1\) and \(t_{2}\), with death counts \(d_1\) and \(d_{2}\) and risk sets \(r_1\) and \(r_{2}=r_1-d_1\)\@.  The contribution to the NLL from these two time points is
\begin{align}
\begin{aligned}
-\log \mathcal{L}_{\text{binomial},1,2} =&-\left[\log\binom{r_1}{d_1} + d_1 \log \left(1-p_1^\text{s}\right) + (r_1-d_1) \log p_1^\text{s}\right] \\
&-\left[\log\binom{r_{1}-d_{1}}{d_{2}} + d_{2} \log \left(1-p_{2}^\text{s}\right) + (r_{1}-d_{1}-d_{2}) \log p_{2}^\text{s}\right].
\end{aligned}
\label{eq:two-timepoints-nll}
\end{align}

The key observation is that \(p_1^\text{s}\) and \(p_2^\text{s}\) only enter the NLL in this term and in the constraint \cref{eq:km-survival}, which only depends on their product \(p_{1,2}^\text{s}=p_1^\text{s} p_2^\text{s}\)\@. We can therefore minimize \cref{eq:two-timepoints-nll} over \(p_1^\text{s}\) and \(p_2^\text{s}\) with the constraint \(p_{1,2}^\text{s}=p_1^\text{s} p_2^\text{s}\)\@.  The result is
\begin{align}
p_1^\text{s} &= \frac{d_1 p_{1,2}^\text{s}+d_2}{d_1+d_2}, \\
p_2^\text{s} &= \frac{(d_1+d_2) p_{1,2}^\text{s}}{d_1 p_{1,2}^\text{s} + d_2}.
\end{align}

We plug into \cref{eq:two-timepoints-nll} to obtain:
\begin{align}
\begin{aligned}
-\log \mathcal{L}_{\text{binomial},1,2} =&-\left[\log\binom{r_1}{d_1+d_2} + (d_1+d_2) \log \left(1-p_{1,2}^\text{s}\right) + (r_1-d_1-d_2) \log p_{1,2}^\text{s}\right] \\
&-\left[\log\binom{d_1+d_2}{d_1} + d_1 \log \left(\frac{d_1}{d_1+d_2}\right) + d_2 \log \left(\frac{d_2}{d_1+d_2}\right)\right],
\end{aligned}
\label{eq:two-timepoints-collapsed-nll}
\end{align}
making use of the binomial coefficient identity
\begin{equation}
\binom{r_1}{d_1}\binom{r_1-d_1}{d_2} = \binom{r_1}{d_1+d_2}\binom{d_1+d_2}{d_1}.
\end{equation}

The first line of \cref{eq:two-timepoints-collapsed-nll} is identical to the NLL contribution from a single time point with \(d_1+d_2\) deaths and risk set \(r_1\)\@. The second line does not depend on \(p_{1,2}^\text{s}\).

This procedure can be applied recursively to collapse any number of consecutive time points with no censoring in between. The result is a single binomial term for the collapsed time points, plus a multinomial NLL term for the number of patients who died at each of the original time points out of the total number of patients who died in the group:
\begin{align}
\begin{aligned}
-\log \mathcal{L}_{\text{binomial},1 \cdots n} =&-\left[\log\binom{r_1}{d_{1 \cdots n}} + d_{1 \cdots n} \log \left(1-p_{1 \cdots n}^\text{s}\right) + (r_1-d_{1 \cdots n}) \log p_{1 \cdots n}^\text{s}\right] \\
&-\left[\log{(d_{1 \cdots n}!)}-d_{1 \cdots n}\log d_{1 \cdots n} - \left(\sum_{i=1}^n \left[\log{(d_i!)} - d_i \log d_i\right]\right)\right],
\end{aligned}
\end{align}
with \(d_{1 \cdots n} = d_1 + d_2 + \cdots + d_n\) and \(p_{1 \cdots n}^\text{s} = p_1^\text{s} p_2^\text{s} \cdots p_n^\text{s}\)\@.

\pkg{KoMbine} automatically collapses such time points before performing the NLL minimization. When applying binomial-only errors, the second line only contributes a constant offset to the NLL\@. Although it does affect the NLL when patient-wise errors are included, it can be precomputed for the various possible values of \(d_i\) and implemented in \pkg{Gurobi} using indicator variables, thereby reducing the number of continuous variables. This can significantly speed up the NLL minimization, especially when not too many patients are censored.

\section{Examples and discussion}

This section provides examples that illustrate interesting aspects of the \KM{} curve estimation.

\begin{figure}
  \centering
  \includegraphics[width=0.5\textwidth]{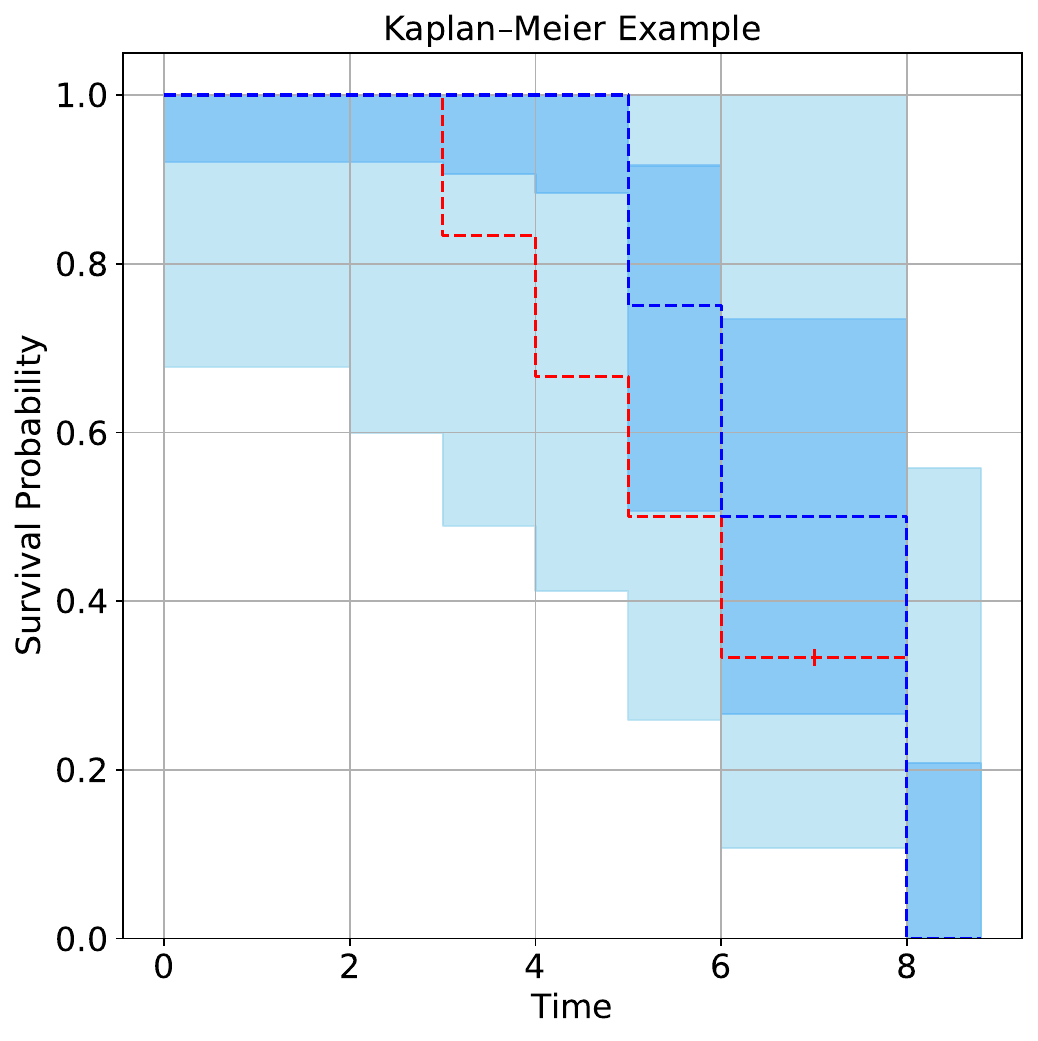}
  \caption{\label{fig:best-fit-vs-nominal} An example \KM{} curve. The best fit (blue) does not coincide with the nominal curve (red). The error bands are shown as shaded blue areas.}
\end{figure}

\subsection%
[Best fit vs. nominal \KM{} curve]
{Best fit vs.\ nominal \KM{} curve}

Counter-intuitively, the best fit \KM{} curve does not necessarily coincide with the nominal \KM{} curve. An example is shown in \cref{fig:best-fit-vs-nominal}.

If only the binomial penalty is included, the survival NLL always minimizes at the nominal survival probability. Similarly, with only the patient-wise errors, the nominal survival probability is also always optimal. However, when both are included, the two may differ.

To understand this, imagine a case where 7 patients were at risk at a particular time point and 4 of them survived. An additional patient who survived is nominally excluded from the curve, but their parameter is so close to the boundary that they can be included with only a small NLL penalty of \(0.01\)\@. \Cref{eq:binomial-nll} evaluates to \(-1.255\) for \(4\) patients surviving out of \(7\) with \(p_\text{s}=4/7\), but to \(-1.267\) for \(5\) patients surviving out of \(8\) with \(p_\text{s}=5/8\). This small difference in the binomial penalty outweighs the even smaller penalty for including the patient, and so the minimum NLL occurs at the best fit survival probability of \(5/8\).

This surprising result highlights the importance of the patient-wise uncertainties. If the NLL penalty for including the patient really is that small, then we cannot, with any confidence, assign the patient to one \KM{} curve or the other.

\subsection{Comparison with conventional methods}\label{sec:compare-to-conventional}

\subsubsection{Greenwood confidence intervals}\label{sec:compare-to-greenwood}

\begin{figure}
\centering
\begin{subfigure}[t]{0.49\textwidth}
  \centering
  \includegraphics[width=\linewidth]{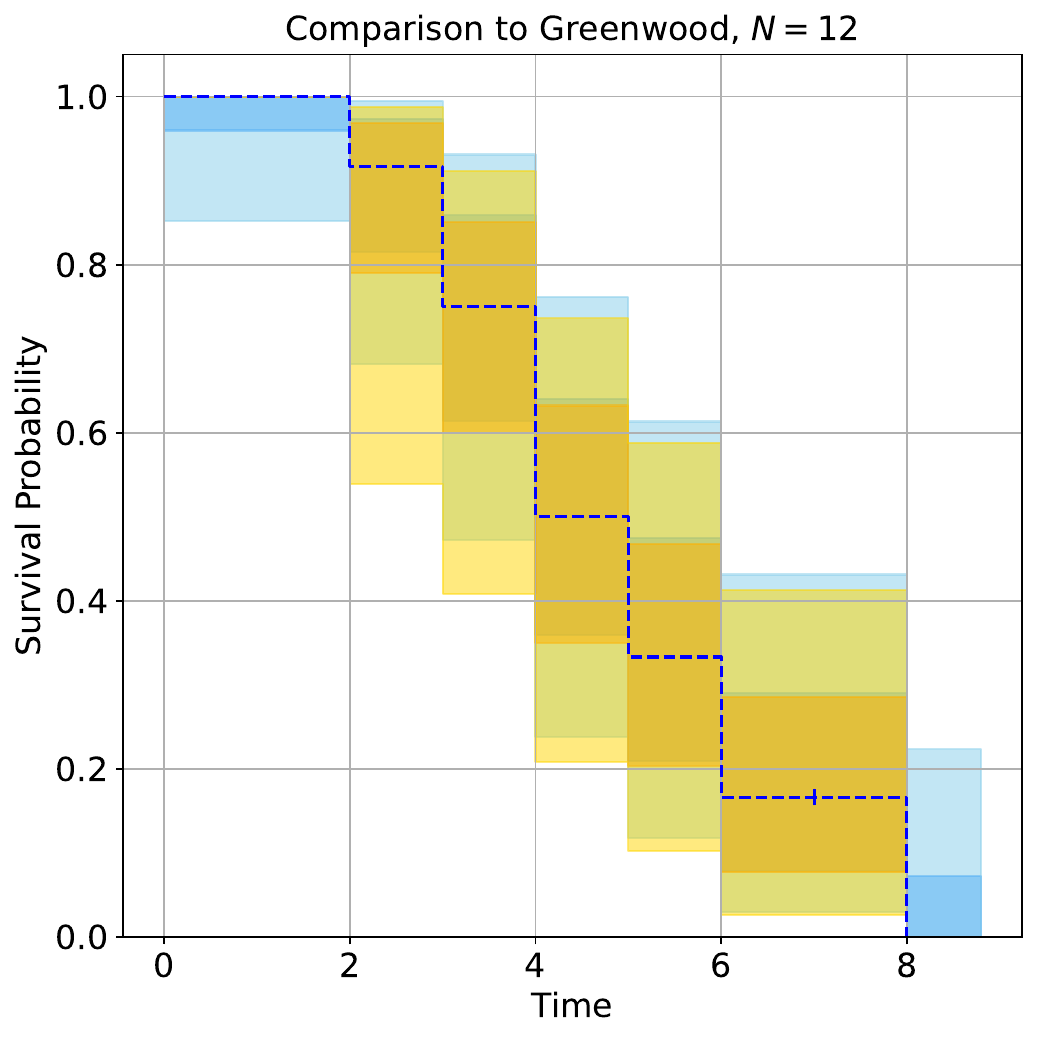}
  \caption{\label{fig:compare-to-greenwood-small-n}}
\end{subfigure}
\begin{subfigure}[t]{0.49\textwidth}
  \centering
  \includegraphics[width=\linewidth]{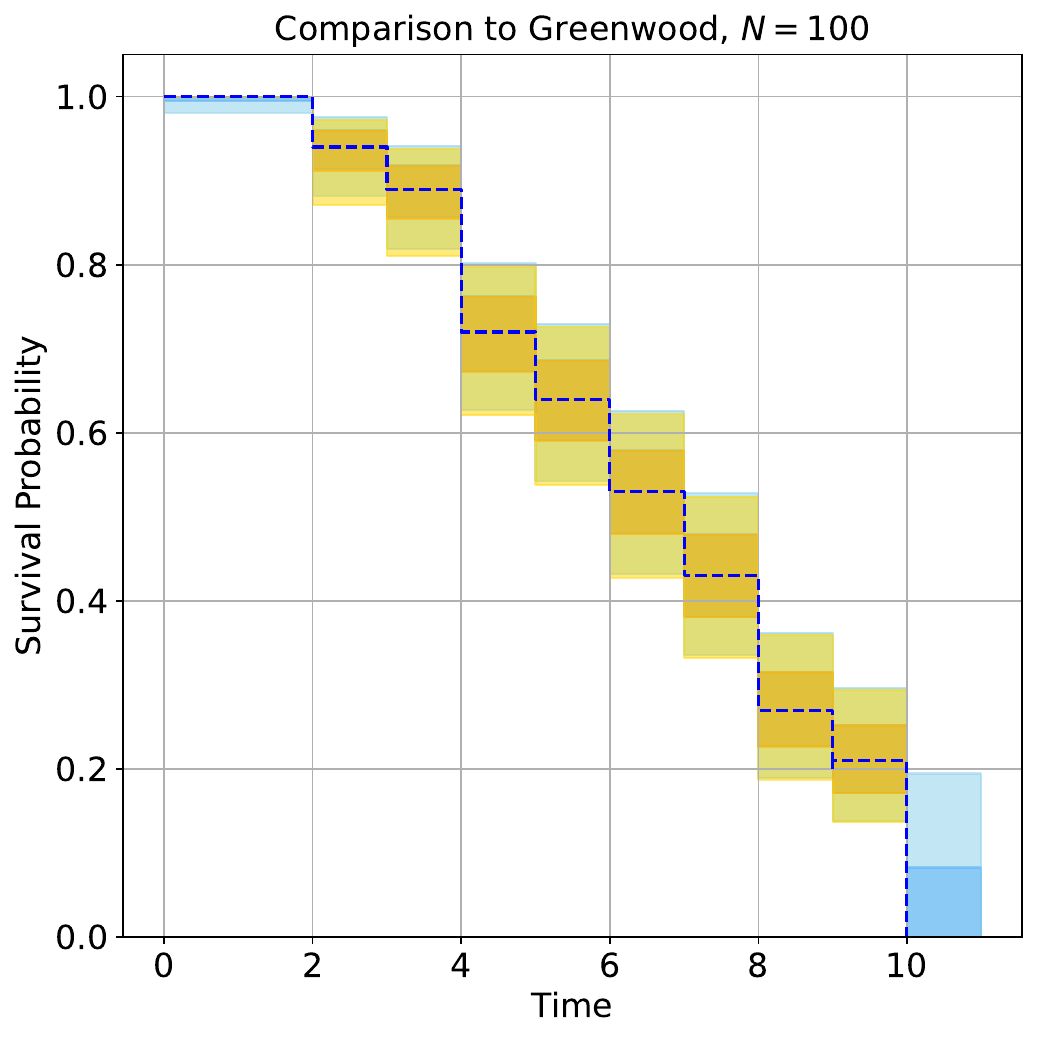}
  \caption{\label{fig:compare-to-greenwood-large-n}}
\end{subfigure}
\caption{\label{fig:compare-to-greenwood} \KM{} curves with confidence intervals using only the binomial error, calculated using the exponential Greenwood (e. G.) formula from \cref{eq:exponential-greenwood} (gold) and the likelihood method from \pkg{KoMbine} (blue). The left panel shows a small cohort of 12 patients, while the right panel shows a larger cohort of 100 patients.}
\end{figure}

\Cref{fig:compare-to-greenwood} shows \KM{} curves with the \pkg{KoMbine} confidence intervals and the exponential Greenwood confidence intervals from \cref{eq:exponential-greenwood}. The \pkg{KoMbine} confidence intervals are calculated with the binomial error only, to facilitate the comparison.

In \cref{fig:compare-to-greenwood-small-n}, only 12 patients are included in the \KM{} curve, and the exponential Greenwood confidence intervals are noticeably different from the \pkg{KoMbine} intervals. \Cref{fig:compare-to-greenwood-large-n} shows \KM{} curves with 100 patients, and the two confidence intervals are in agreement.

One striking difference in both cases is that the Greenwood method cannot calculate confidence intervals for the first or last time points, when the probabilities are \(1\) and \(0\). This is a fundamental limitation of the Greenwood method, which approximates the uncertainty on the \emph{true} binomial probability using the \emph{observed} numbers of patients who were at risk and who survived. If everyone survived or died, the binomial uncertainty is \(0\).

Our method, by contrast, calculates the binomial likelihood (and hence the uncertainty) using the \emph{true} survival probability, which we can do despite not knowing it in advance, because it is a parameter in the likelihood fit. This is the fundamental tradeoff of the likelihood method: we can be more precise, but the likelihood fit is more computationally intensive.

\subsubsection[Conventional log-rank p~value estimation]{Conventional log-rank \(p\)~value estimation}\label{sec:compare-to-conventional-p-value}

\begin{figure}[ht]
\centering
\begin{subfigure}[t]{0.32\textwidth}
  \centering
  \includegraphics[width=\linewidth]{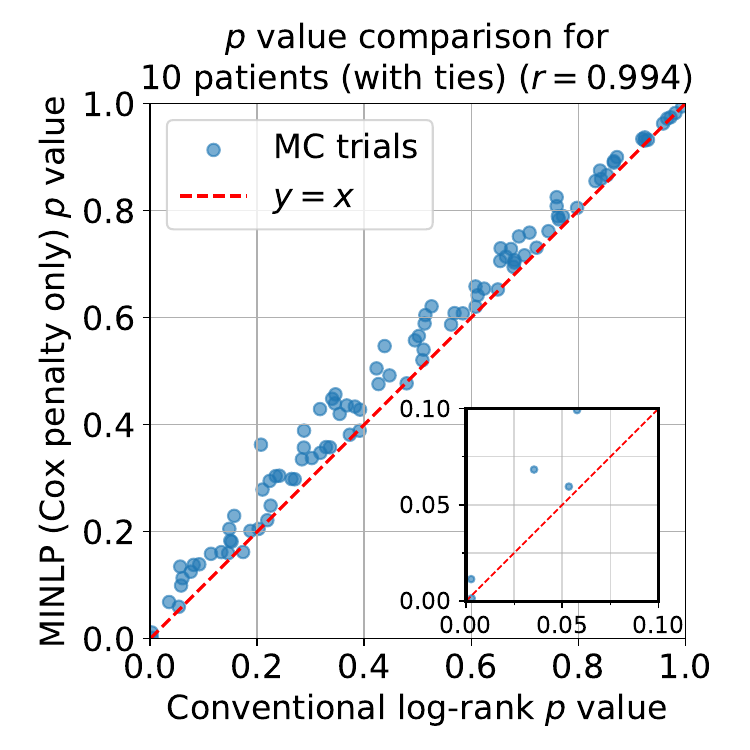}
  \caption{\label{fig:compare-p-value-10-patients}}
\end{subfigure}
\begin{subfigure}[t]{0.32\textwidth}
  \centering
  \includegraphics[width=\linewidth]{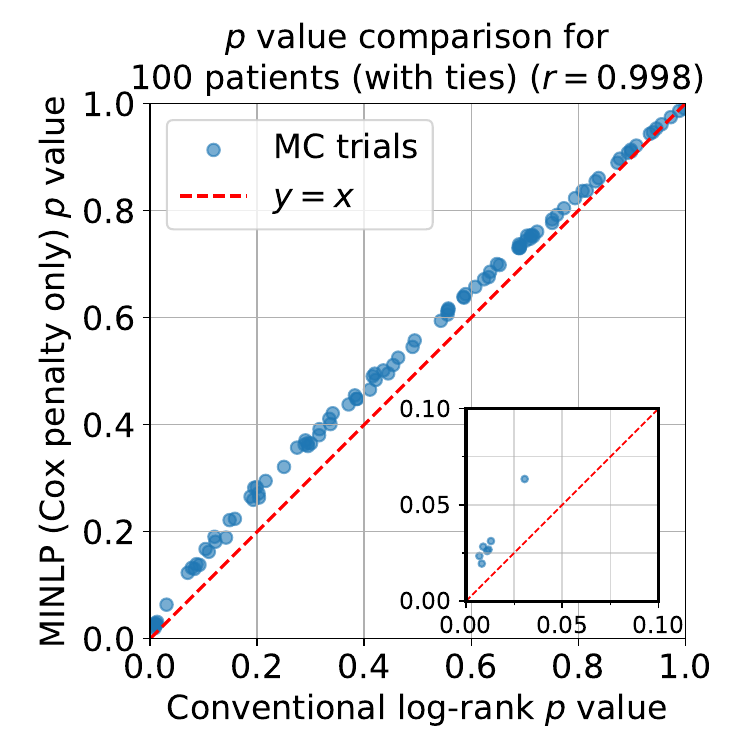}
  \caption{\label{fig:compare-p-value-100-patients}}
\end{subfigure}
\begin{subfigure}[t]{0.32\textwidth}
  \centering
  \includegraphics[width=\linewidth]{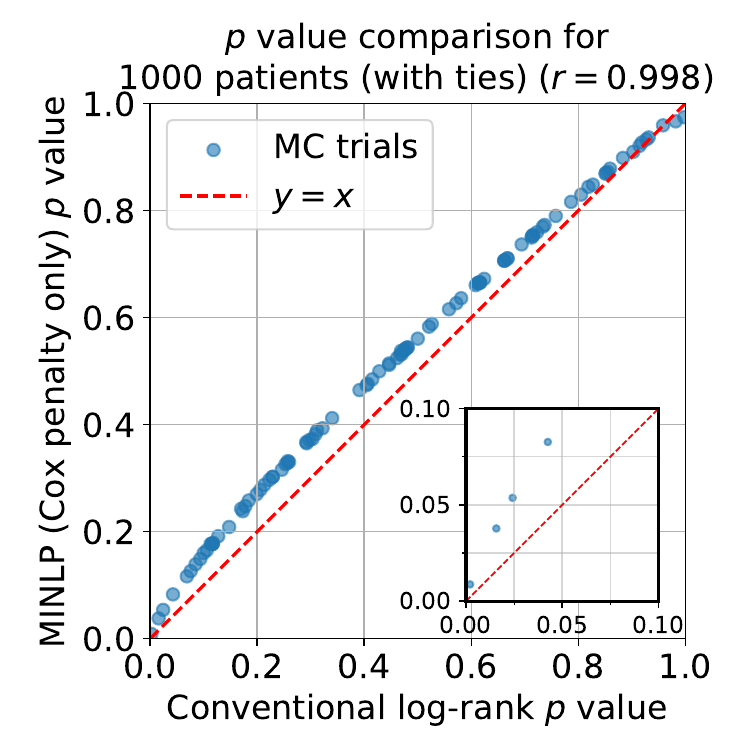}
  \caption{\label{fig:compare-p-value-1000-patients}}
\end{subfigure}
\begin{subfigure}[t]{0.32\textwidth}
  \centering
  \includegraphics[width=\linewidth]{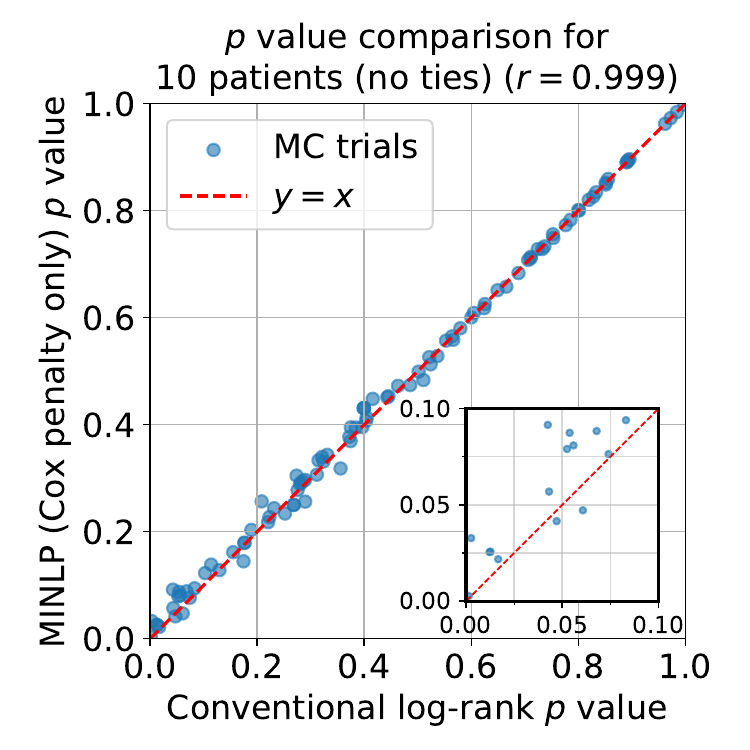}
  \caption{\label{fig:compare-p-value-10-patients-no-ties}}
\end{subfigure}
\begin{subfigure}[t]{0.32\textwidth}
  \centering
  \includegraphics[width=\linewidth]{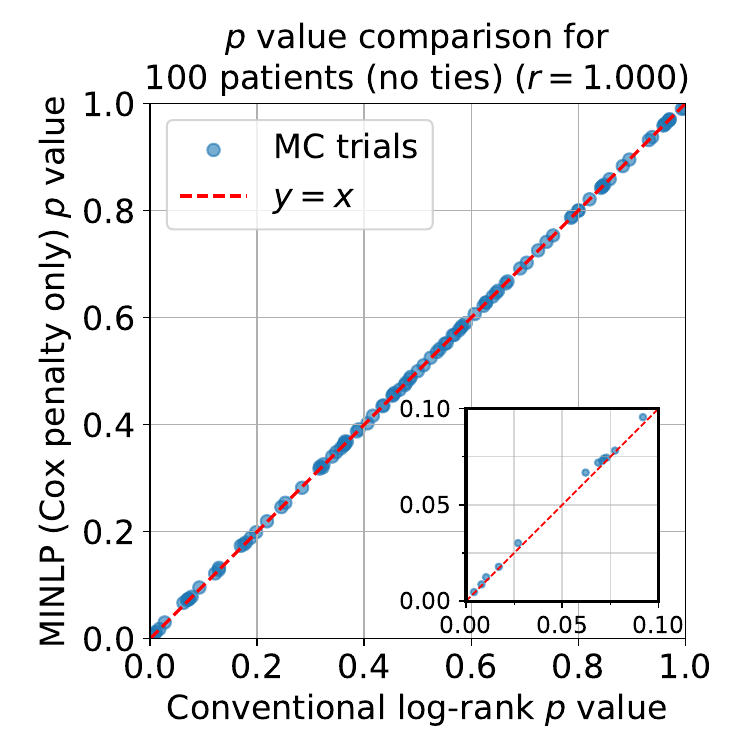}
  \caption{\label{fig:compare-p-value-100-patients-no-ties}}
\end{subfigure}
\begin{subfigure}[t]{0.32\textwidth}
  \centering
  \mbox{} 
\end{subfigure}
\caption{\label{fig:compare-p-value} Scatter plot showing the \(p\)~values computed using the conventional log-rank method (horizontal axis) and the likelihood method using only the Cox penalty (vertical axis). Each point represents a trial with randomly simulated patients. The left, middle, and right columns show cohorts of 10, 100, and 1000 patients, respectively. The top row allows patients to die at the same time, while the bottom row does not.}
\end{figure}

In order to compare the \(p\)~values, we generate Monte Carlo cohorts of patients. We calculate the \(p\)~values using our method, disabling the patient-wise penalties, and with the conventional log-rank test. We plot the results in \cref{fig:compare-p-value}. The top row allows patients to die at the same time, which means that we rely on the Breslow approximation in \cref{eq:cox-nll-breslow}. In the bottom row, no two patients die at the same time, so the Breslow approximation is identical to the exact formula in \cref{eq:cox-nll-exact}.

When there are no ties, the points scatter along the diagonal line \(y=x\), meaning that, on average, the two methods give identical results. When ties are allowed, the diagonal line is slightly altered to a curve. The Breslow approximation gives more conservative \(p\)~values, and the effect is most pronounced for small \(p\)~values.

The Monte Carlo patients for the top row are simulated with integer survival times from \(1\) to \(10\), which means that we expect a tie rate of about \(10\%\). The error introduced by the Breslow approximation will increase with a higher tie rate.

The scatter from the line or curve, which shrinks with larger cohort sizes, is because the full likelihood method avoids the asymptotic approximations used in the conventional method, similar to the difference between the \pkg{KoMbine} and Greenwood confidence intervals in \cref{sec:compare-to-greenwood}.

\subsection{Lung cancer dataset}

\begin{figure}[ht]
  \newcommand{\figwidth}{0.28\textwidth}
  \newcommand{\legendwidth}{0.13\textwidth}
  \newcommand{\movelegendup}{\vspace{-0.9cm}}
  \centering
  \begin{subfigure}[c]{\figwidth}
    \centering
    \includegraphics[width=\linewidth]{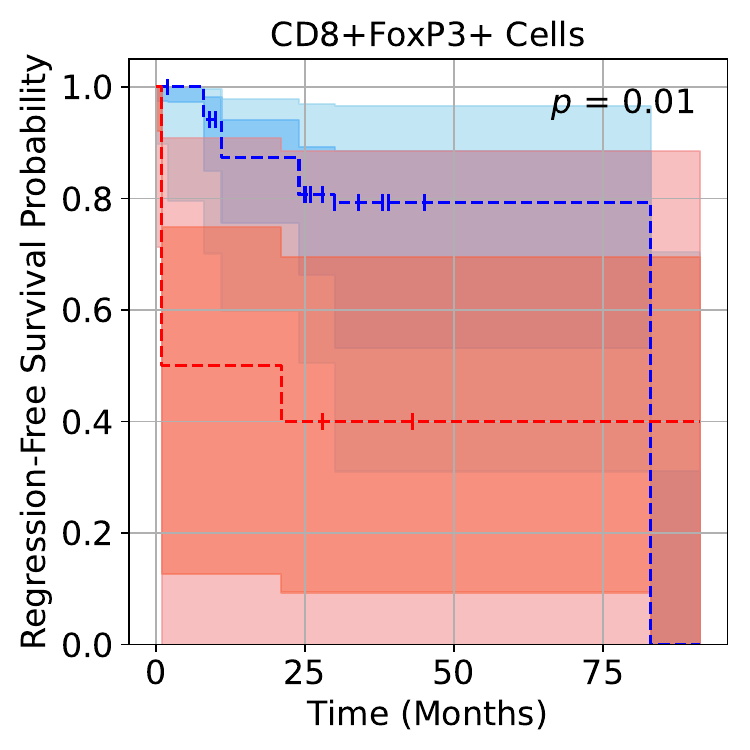}
    \caption{\label{fig:lung-dataset-cells}}
  \end{subfigure}
  \begin{subfigure}[c]{\figwidth}
    \centering
    \includegraphics[width=\linewidth]{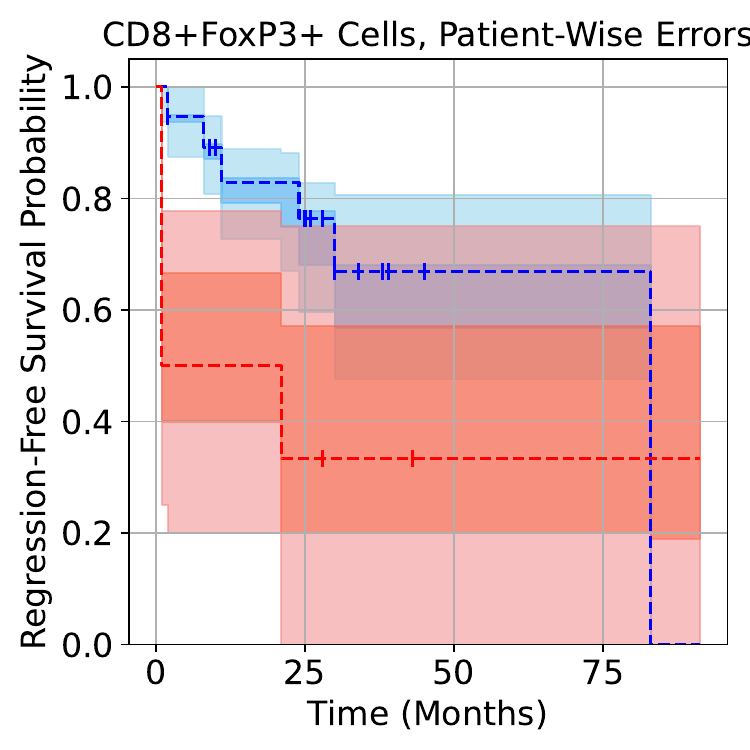}
    \caption{\label{fig:lung-dataset-cells-patient-wise}}
  \end{subfigure}
  \begin{subfigure}[c]{\figwidth}
    \centering
    \includegraphics[width=\linewidth]{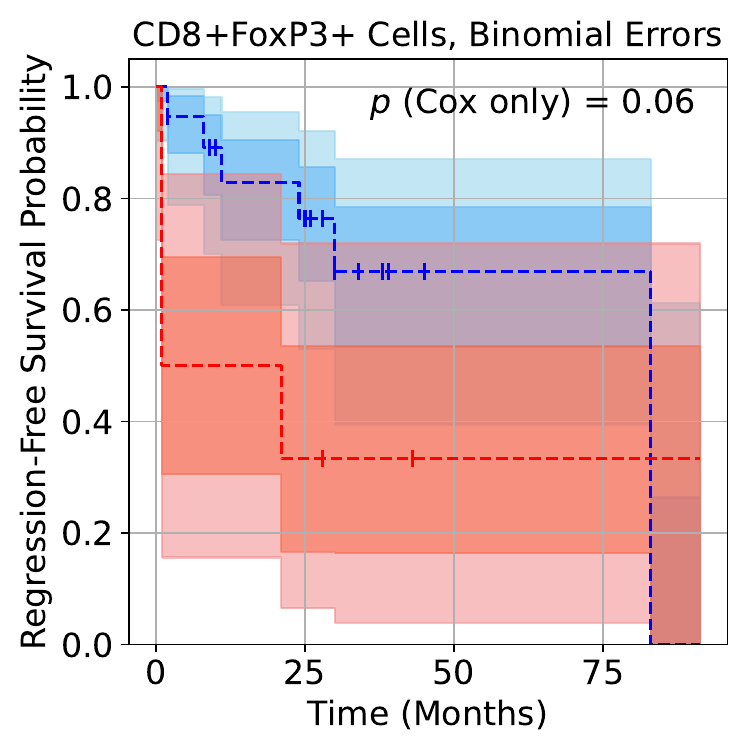}
    \caption{\label{fig:lung-dataset-cells-binomial}}
  \end{subfigure}
  \begin{minipage}[c]{\legendwidth}
    \centering
    \movelegendup
    \includegraphics[width=\linewidth]{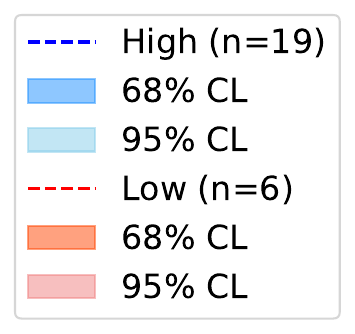}
  \end{minipage}
  \begin{subfigure}[c]{\figwidth}
    \centering
    \includegraphics[width=\linewidth]{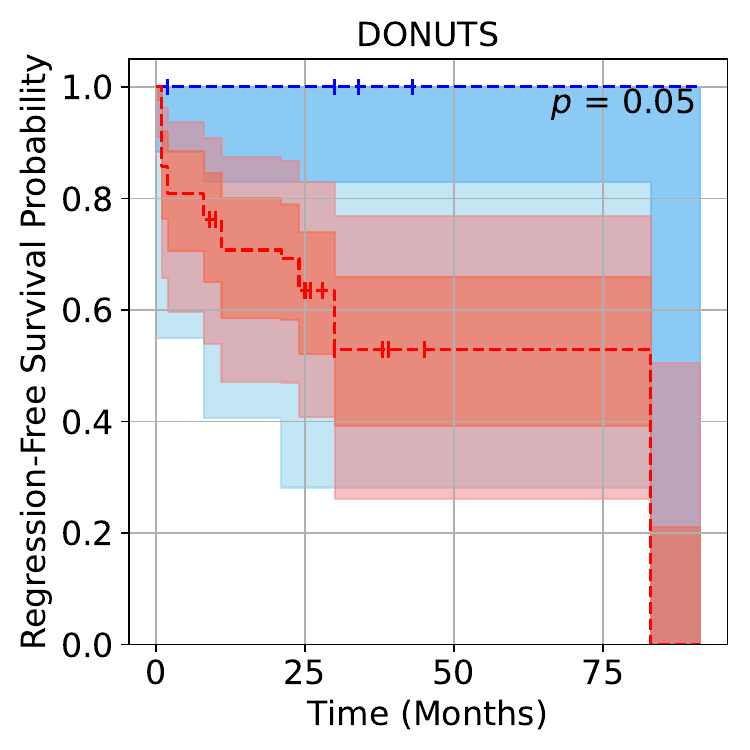}
    \caption{\label{fig:lung-dataset-donuts}}
  \end{subfigure}
  \begin{subfigure}[c]{\figwidth}
    \centering
    \includegraphics[width=\linewidth]{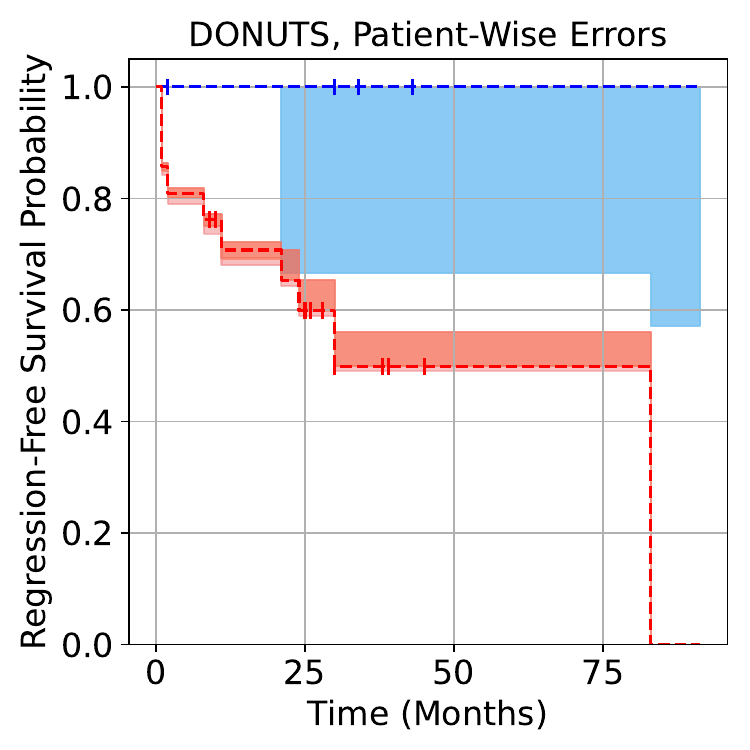}
    \caption{\label{fig:lung-dataset-donuts-patient-wise}}
  \end{subfigure}
  \begin{subfigure}[c]{\figwidth}
    \centering
    \includegraphics[width=\linewidth]{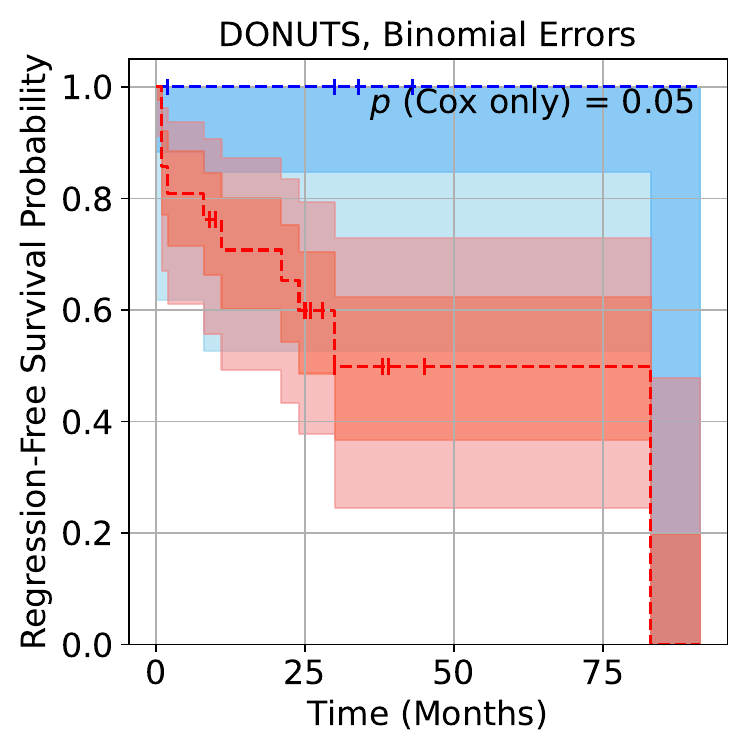}
    \caption{\label{fig:lung-dataset-donuts-binomial}}
  \end{subfigure}
  \begin{minipage}[c]{\legendwidth}
    \centering
    \movelegendup
    \includegraphics[width=\linewidth]{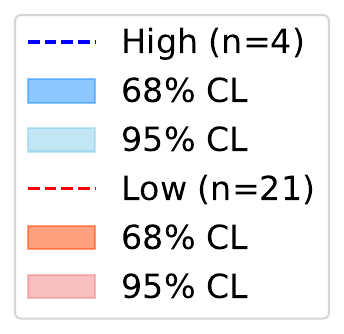}
  \end{minipage}
  \caption{\label{fig:lung-dataset} \KM{} curves for the regression-free survival of patients with lung cancer who were treated with neoadjuvant anti-PD-1 immunotherapy, stratifying patients by their pre-treatment densities of either CD8+FoxP3+ cells (top) or DONUTS (bottom). The left column shows the full uncertainty, while the middle and right columns show the uncertainties broken down into patient-wise and binomial contributions, respectively.}
\end{figure}

This example is taken from \citet{DONUTS}, in which we analyzed pre-treatment biopsies from a cohort of patients with non-small-cell lung cancer (NSCLC) treated with neoadjuvant anti-PD-1 immunotherapy. We identified CD8+FoxP3+ cells, which are associated with good outcomes, but are extremely rare: a patient may have only one or two in their biopsy. We also developed a probabilistic biomarker, called DONUTS, which resemble the niche of the CD8+FoxP3+ cells but are much more abundant.

\Cref{fig:lung-dataset-cells} shows \KM{} curves for regression-free survival (RFS) of patients with high and low densities of CD8+FoxP3+ cells. The patients with a high density of CD8+FoxP3+ cells have better outcomes, as expected, but the error bands are large due to the high Poisson uncertainty in the number of CD8+FoxP3+ cells. By contrast, \Cref{fig:lung-dataset-donuts} shows the \KM{} curves for RFS of patients with high and low densities of DONUTS\@. The error bands are smaller, as the DONUTS are much more abundant.

\Cref{fig:lung-dataset-cells-patient-wise,fig:lung-dataset-donuts-patient-wise,fig:lung-dataset-cells-binomial,fig:lung-dataset-donuts-binomial} show the same \KM{} curves with the uncertainties broken down into the patient-wise and binomial contributions. The binomial uncertainties have similar magnitude for the cells and for the DONUTS, as both have similar sample sizes in the high and low categories. The patient-wise uncertainties, on the other hand, are comparable to the binomial errors for the cells but almost zero for the DONUTS\@.

It should be noted that this is a fairly small cohort of only 25 patients. In a larger-scale study, the binomial uncertainties would be smaller and the difference in patient-wise uncertainties would be even more pronounced.

\section{Future work}

\subsection{Correlated uncertainties}

One important feature of \pkg{Combine} is uncertainties that are correlated between different channels. For example, uncertainties on the electron resolution affect, in different ways, the Higgs boson decay to four electrons or to two electrons and two muons.

Similarly, uncertainties resulting from batch effects in a biomarker assay will affect all patients in the same batch\@. \pkg{KoMbine} does not currently support this correlation. To implement it, \cref{eq:patient-nll} would need to be modified, as there would no longer be a fixed penalty per patient, but rather a penalty that depends on the other patients in the same batch. Furthermore, if there are multiple uncertainties correlated between different groups of patients, the penalty could depend on all patients at once.

The current \pkg{Gurobi} model would no longer work, as it requires precomputing \(\mathcal{L}_j^{\text{patient}}\). With \(N\) patients, we can easily precompute \(N\) values, but not \(2^N\). Instead, we would likely need to pass the uncertainty parameters \(\theta_k\) directly to the \pkg{Gurobi} model and let it compute \(a_j\left(\theta_k\right)\) and hence \(\mathcal{L}_{\text{patient}}\left(\theta_k\right)\) internally.

The increased number of continuous variables will probably require a rethinking of how the model is solved\@. \pkg{Gurobi} is not optimized for the types of complex nonlinear relationships that would be necessary here\@. \pkg{Minuit}, which forms the backbone for \pkg{Combine}, \emph{is} optimized for many continuous nuisance variables, but cannot handle indicator variables such as our \(a_j\), which are fundamental to the logic of \pkg{KoMbine}. Combining the strengths of both will be an interesting and nontrivial challenge.

\subsection{More advanced survival functionality}

The \pkg{lifelines} package provides support for other types of survival analysis, such as parameteric survival models, left and interval censoring, and other types of visualizations, such as cumulative hazard plots\@. \pkg{KoMbine} currently only supports the basic \KM{} curve, including right censoring, but similar statistical methods could be used to incorporate these other features.

\subsection{ROC curves}

Receiver operating characteristic (ROC) curves are also commonly used to evaluate biomarkers and predict patient outcomes. Similar statistical methods are necessary to propagate uncertainties through the ROC curve estimation process.

Some preliminary work, called \pkg{ROC Picker}, is available in the \pkg{KoMbine} package, but it is not yet fully functional. Crucially, it can handle either binomial or patient-wise uncertainties, but not both at the same time. ROC curve estimation is more complex because the uncertainties are correlated across all points in the curve, and the metrics typically used, such as the area under the curve (AUC), are nontrivial functions of the entire curve.

\section{Conclusion}

\pkg{KoMbine} provides a new method to propagate statistical and systematic uncertainties through the \KM{} curve estimation process. The handling of the binomial uncertainty is more precise than the Greenwood confidence intervals used in other packages, and the patient-wise uncertainties are implemented for the first time.

This package will enable uncertainty estimates on \KM{} curves to take \emph{all} sources of uncertainty into account, facilitating more robust and reproducible biomarker studies. Furthermore, by allowing the user to plot the different uncertainties separately, it will help researchers understand how to improve the analysis, furthering the development of the next generation of biomarkers.

\section*{Computational details}

The results in this paper were obtained using \proglang{Python}~3.11.13 with \pkg{numpy}~2.3.2, \pkg{scipy}~1.16.1, and \pkg{gurobipy}~12.0.3, as well as \pkg{Gurobi}~10.0.3. All \proglang{Python} packages can be installed using \pkg{pip} or \pkg{conda}.

\pkg{Gurobi} can be downloaded from \url{https://www.gurobi.com/}. Although it requires a license, free licenses are available for academic use.

\section*{Acknowledgments}

I would like to thank my AstroPath colleagues, in particular my mentors, Drs.~Alex Szalay and Janis Taube, for their support on this project. I would also like to acknowledge editing support by Ben Cohen, provided by the Office of the Vice Provost for Research at Johns Hopkins University.

This research was supported by the Mark Foundation for Cancer Research.


\bibliography{02_kombine}

\begin{thebibliography}{19}
\newcommand{\enquote}[1]{``#1''}
\providecommand{\natexlab}[1]{#1}
\providecommand{\url}[1]{\texttt{#1}}
\providecommand{\urlprefix}{URL }
\expandafter\ifx\csname urlstyle\endcsname\relax
  \providecommand{\doi}[1]{doi:\discretionary{}{}{}#1}\else
  \providecommand{\doi}{doi:\discretionary{}{}{}\begingroup
  \urlstyle{rm}\Url}\fi
\providecommand{\eprint}[2][]{\url{#2}}

\bibitem[{Brent(1973)}]{brentq}
Brent RP (1973).
\newblock \emph{Algorithms for Minimization without Derivatives}.
\newblock Courier Corporation.

\bibitem[{Breslow(1974)}]{breslow1974covariance}
Breslow N (1974).
\newblock \enquote{Covariance Analysis of Censored Survival Data.}
\newblock \emph{Biometrics}, \textbf{30}(1), 89--99.
\newblock \doi{10.2307/2529620}.

\bibitem[{Chatrchyan \emph{et~al.}(2012)}]{HIG-12-028}
Chatrchyan S, \emph{et~al.} (CMS) (2012).
\newblock \enquote{Observation of a New Boson at a Mass of 125 {GeV} with the
  {CMS} Experiment at the {LHC}.}
\newblock \emph{Physics Letters B}, \textbf{716}, 30--61.
\newblock \doi{10.1016/j.physletb.2012.08.021}.
\newblock \eprint{arXiv:1207.7235 [hep-ex]}.

\bibitem[{Cottrell \emph{et~al.}(2025)Cottrell, Roskes, Fotheringham, Cohen,
  Zhang, Engle, Wang, Will, Sunshine, Jim\'enez-S\'anchez, Zeng, Caushi, Zhang,
  D'amiano, Deutsch, Uttam, Pirie, Vlaminck, Mataj, Fiorante, Espinosa, Popa,
  Ogurtsova, Soto-Diaz, Eminizer, Tabrisky, Jorquera, Skidmore, Medvedev,
  Chaft, Brahmer, Conroy, Reuss, Danilova, Ji, Forde, Pardoll, Smith, Green,
  Szalay, and Taube}]{DONUTS}
Cottrell TR, Roskes JS, Fotheringham M, Cohen E, Zhang B, Engle LL, Wang D,
  Will E, Sunshine JC, Jim\'enez-S\'anchez D, Zeng Z, Caushi JX, Zhang J,
  D'amiano NM, Deutsch JS, Uttam S, Pirie K, Vlaminck D, Mataj M, Fiorante A,
  Espinosa N, Popa T, Ogurtsova A, Soto-Diaz S, Eminizer M, Tabrisky S,
  Jorquera A, Skidmore J, Medvedev D, Chaft JE, Brahmer JR, Conroy M, Reuss JE,
  Danilova L, Ji H, Forde PM, Pardoll DM, Smith KN, Green BF, Szalay AS, Taube
  JM (2025).
\newblock \enquote{Novel Predictive Spatial Biomarker in Non-Small Cell Lung
  Carcinoma: The Diversity of Niches Unlocking Treatment Sensitivity (DONUTS).}
\newblock Submitted to \emph{Nature}.

\bibitem[{Cox(1972)}]{cox1972regression}
Cox DR (1972).
\newblock \enquote{Regression Models and Life-Tables.}
\newblock \emph{Journal of the Royal Statistical Society: Series B
  (Methodological)}, \textbf{34}(2), 187--220.
\newblock \doi{10.1111/j.2517-6161.1972.tb00899.x}.

\bibitem[{Davidson-Pilon(2019)}]{lifelines}
Davidson-Pilon C (2019).
\newblock \enquote{\pkg{lifelines}: Survival Analysis in \proglang{Python}.}
\newblock \emph{Journal of Open Source Software}, \textbf{4}(40), 1317.
\newblock \doi{10.21105/joss.01317}.

\bibitem[{Greenwood(1926)}]{Greenwood}
Greenwood M (1926).
\newblock \enquote{The Natural Duration of Cancer.}
\newblock \emph{Reports on Public Health and Medical Subjects}, \textbf{33},
  1--26.

\bibitem[{{\pkg{Gurobi} Optimization, LLC}(2024)}]{gurobi}
{\pkg{Gurobi} Optimization, LLC} (2024).
\newblock \enquote{\pkg{Gurobi} Optimizer Reference Manual.}
\newblock \urlprefix\url{https://www.gurobi.com}.

\bibitem[{Hayrapetyan \emph{et~al.}(2024)}]{CAT-23-001}
Hayrapetyan A, \emph{et~al.} (CMS) (2024).
\newblock \enquote{The {CMS} Statistical Analysis and Combination Tool:
  \pkg{Combine}.}
\newblock \emph{Computing and Software for Big Science}, \textbf{8}(1), 19.
\newblock \doi{10.1007/s41781-024-00121-4}.
\newblock \eprint{arXiv:2404.06614 [physics.data-an]}.

\bibitem[{Hosmer \emph{et~al.}(2008)Hosmer, Lemeshow, and
  May}]{appliedsurvivalanalysis}
Hosmer DW, Lemeshow S, May S (2008).
\newblock \emph{Applied Survival Analysis: Regression Modeling of Time-to-Event
  Data}.
\newblock 2nd edition. John Wiley \& Sons, Hoboken, NJ.
\newblock ISBN 978-0-470-07963-5.
\newblock \doi{10.1002/9780470258019}.

\bibitem[{James(1994)}]{minuit}
James F (1994).
\newblock \enquote{\pkg{MINUIT} Function Minimization and Error Analysis:
  Reference Manual Version 94.1.}

\bibitem[{Mantel(1966)}]{mantel1966evaluation}
Mantel N (1966).
\newblock \enquote{Evaluation of Survival Data and Two New Rank Order
  Statistics Arising in Its Consideration.}
\newblock \emph{Cancer Chemotherapy Reports}, \textbf{50}(3), 163--170.
\newblock PMID: 5910392.

\bibitem[{Mehta(1991)}]{StatXact}
Mehta CR (1991).
\newblock \enquote{\pkg{StatXact}: A Statistical Package for Exact
  Nonparametric Inference.}
\newblock \emph{The American Statistician}, \textbf{45}(1), 74--75.
\newblock \doi{10.2307/2685246}.

\bibitem[{Peto and Peto(1972)}]{peto1972asymptotically}
Peto R, Peto J (1972).
\newblock \enquote{Asymptotically Efficient Rank Invariant Test Procedures.}
\newblock \emph{Journal of the Royal Statistical Society. Series A (General)},
  \textbf{135}(2), 185--207.
\newblock \doi{10.2307/2344317}.

\bibitem[{{\proglang{R} Core Team}(2023)}]{R}
{\proglang{R} Core Team} (2023).
\newblock \emph{\proglang{R}: A Language and Environment for Statistical
  Computing}.
\newblock \proglang{R} Foundation for Statistical Computing, Vienna, Austria.
\newblock \urlprefix\url{https://www.R-project.org/}.

\bibitem[{Sawyer(2003)}]{GreenwoodNotes}
Sawyer S (2003).
\newblock \enquote{The {G}reenwood and Exponential {G}reenwood Confidence
  Intervals in Survival Analysis.}
\newblock Unpublished notes,
  \urlprefix\url{https://www.math.wustl.edu/~sawyer/handouts/greenwood.pdf}.

\bibitem[{Therneau(2024)}]{survival-package}
Therneau TM (2024).
\newblock \emph{A Package for Survival Analysis in \proglang{R}}.
\newblock \proglang{R} package version 3.8-3,
  \urlprefix\url{https://CRAN.R-project.org/package=survival}.

\bibitem[{Verkerke and Kirkby(2003)}]{RooFit}
Verkerke W, Kirkby DP (2003).
\newblock \enquote{{The \pkg{RooFit} Toolkit for Data Modeling}.}
\newblock \eprint{arXiv:physics/0306116}.

\bibitem[{Virtanen \emph{et~al.}(2020)Virtanen, Gommers, Oliphant, Haberland,
  Reddy, Cournapeau, Burovski, Peterson, Weckesser, Bright, {van der Walt},
  Brett, Wilson, Millman, Mayorov, Nelson, Jones, Kern, Larson, Carey, Polat,
  Feng, Moore, {VanderPlas}, Laxalde, Perktold, Cimrman, Henriksen, Quintero,
  Harris, Archibald, Ribeiro, Pedregosa, {van Mulbregt}, and {\pkg{SciPy} 1.0
  Contributors}}]{scipy}
Virtanen P, Gommers R, Oliphant TE, Haberland M, Reddy T, Cournapeau D,
  Burovski E, Peterson P, Weckesser W, Bright J, {van der Walt} SJ, Brett M,
  Wilson J, Millman KJ, Mayorov N, Nelson ARJ, Jones E, Kern R, Larson E, Carey
  CJ, Polat {\.I}, Feng Y, Moore EW, {VanderPlas} J, Laxalde D, Perktold J,
  Cimrman R, Henriksen I, Quintero EA, Harris CR, Archibald AM, Ribeiro AH,
  Pedregosa F, {van Mulbregt} P, {\pkg{SciPy} 10 Contributors} (2020).
\newblock \enquote{\pkg{SciPy} 1.0: Fundamental Algorithms for Scientific
  Computing in \proglang{Python}.}
\newblock \emph{Nature Methods}, \textbf{17}, 261--272.
\newblock \doi{10.1038/s41592-019-0686-2}.

\end{thebibliography}


\newpage

\begin{appendix}

\section{Datacard format}\label{app:datacard}

The inputs to \pkg{KoMbine} are provided in a datacard format, similar to the one used in \pkg{Combine} \citep{CAT-23-001}. The datacard is a text file with the following format:
\begin{verbatim}
# Datacard for KoMbine
# This is a comment line
observable_type fixed
------------
# List of patients
------------
survival_time 3   4   5   2   4   5   6   4   3   6   8   7
censored      0   0   0   0   0   0   0   0   0   0   0   1
observable    0.1 0.2 0.3 0.4 0.5 0.6 0.3 0.4 0.5 0.6 0.7 0.8
\end{verbatim}
\end{appendix}
Any line starting with \code{\#} is a comment and is ignored, as are lines of hyphens \code{------------}.

The first line is the header, which specifies the type of observable. In this case, the observable is fixed, which means that there are no patient-wise uncertainties. The \code{survival\_time} gives the length of time the patient was observed before being censored or dying, and the \code{censored} line indicates whether the patient was censored (1) or not (0).

The \code{observable} line gives the value of the observable for each patient. The format is similar for other \code{observable\_type}s. This datacard gives the same nominal \KM{} curve as the previous example, but the patient-wise uncertainties are modeled as a ratio of two Poisson distributions:
\begin{verbatim}
observable_type poisson_ratio
------------
# List of patients
------------
survival_time 3   4   5   2   4   5   6   4   3   6   8   7
censored      0   0   0   0   0   0   0   0   0   0   0   1
num           10  20  30  40  50  60  30  40  50  60  70  80
denom         100 100 100 100 100 100 100 100 100 100 100 100
\end{verbatim}

To create a \KM{} plot, you can run:
\begin{CodeInput}
kombine datacard.txt km_plot.pdf --parameter-min 0.45
\end{CodeInput}
This will create a \KM{} plot with the patients whose parameter is at least \(0.45\) included in the curve.

You can also create a \KM{} plot with two curves, for patients with a high and low value of the observable:
\begin{CodeInput}
kombine_twogroups datacard.txt --parameter-threshold 0.45
\end{CodeInput}

The \code{--include-binomial-only} and \code{--include-patient-wise-only} options will calculate error bands that only include the binomial or patient-wise uncertainties, respectively, as described in \cref{sec:isolating-uncertainties}. By default, the error bands will be drawn hatched, with the combined uncertainty also displayed on the plot. If that is too hard to see, you can disable the combined uncertainty with \code{--exclude-full-nll}.

Various additional command line arguments are available and are documented in \code{kombine --help} or \code{kombine_twogroups --help}.

\end{document}